\documentclass[prb,amsmath,amssymb,twocolumn,showpacs,floatfix]{revtex4}
\usepackage{graphicx}
\usepackage{epsfig}
\usepackage{dcolumn}
\usepackage{bm}
\begin{document}

\title{Hole doping dependences of the magnetic penetration depth
and vortex core size in YBa$_2$Cu$_3$O$_y$:
Evidence for stripe correlations near 1/8 hole doping}

\author{J.E.~Sonier,$^{1,2,*}$ S.A.~Sabok-Sayr,$^1$ F.D.~Callaghan,$^1$ 
C.V.~Kaiser,$^1$ V.~Pacradouni,$^1$ J.H.~Brewer,$^{2,3}$ S.L.~Stubbs,$^3$ 
W.N.~Hardy,$^{2,3}$ D.A.~Bonn$^{2,3}$, R.~Liang$^{2,3}$ and W.A.~Atkinson$^4$}

\affiliation{$^1$Department of Physics, Simon Fraser University, Burnaby, British Columbia V5A 1S6, Canada \\
$^2$Canadian Institute for Advanced Research, 180 Dundas Street West, Toronto, Ontario M5G 1Z8, Canada \\
$^3$Department of Physics and Astronomy, University of British Columbia, Vancouver, British Columbia V6T 1Z1, Canada \\
$^4$Department of Physics and Astronomy, Trent University, Peterborough, Ontario K9J 7B8, Canada}

\date{\today}
\begin{abstract}
We report on muon spin rotation ($\mu$SR) measurements of the 
internal magnetic field distribution $n(B)$ in the vortex solid phase 
of YBa$_2$Cu$_3$O$_y$ (YBCO) single crystals, from which we have
simultaneously determined the hole doping dependences of the
in-plane Ginzburg-Landau (GL) length scales in the underdoped
regime. We find that $T_c$ has 
a sublinear dependence on $1/\lambda_{ab}^2$, where $\lambda_{ab}$ 
is the in-plane magnetic penetration depth in the extrapolated limits 
$T \! \rightarrow \! 0$ and $H \! \rightarrow \! 0$. The power 
coefficient of the sublinear dependence is close to that determined 
in severely underdoped YBCO thin films, indicating that the same 
relationship between $T_c$ and the superfluid density is maintained 
throughout the underdoped regime. The GL coherence length $\xi_{ab}$ 
(vortex core size) is found to increase with decreasing hole doping
concentration, and exhibit a field dependence that is explained
by proximity-induced superconductivity on the CuO chains.
Both $\lambda_{ab}$ and $\xi_{ab}$ are enhanced near 1/8 hole doping,
supporting the belief by some that stripe correlations are
a universal property of high-$T_c$ cuprates.
\end{abstract}

\maketitle

\section{Introduction}

Abrikosov vortices in a superconductor are governed by two characteristic 
length scales. The core of a vortex has a size dependent on the 
Ginzburg-Landau (GL) coherence length $\xi$, while the supercurrents 
circulating around the core decay on the scale
of the GL penetration depth $\lambda$. 
In the early days of high-$T_c$ superconductivity it was common practice to 
infer the behavior of the in-plane magnetic penetration depth
$\lambda_{ab}$ from measurements of the 
muon depolarization rate $\sigma$ in the vortex state of polycrystalline samples. 
The temperature dependence of $\sigma$ was found to be consistent with 
$s$-wave pairing symmetry \cite{Harshman:87,Uemura:88,Harshman:89,Pumpin:90} 
and a universal linear scaling of $T_c$ with $\sigma$ was observed in the 
underdoped regime (the so-called `Uemura plot'), indicating that 
$T_c \! \propto \! 1/\lambda_{ab}^2 \! \propto \! \rho_s$,
where $\rho_s$ is the superfluid density.\cite{Uemura:89,Uemura:91}
Later, microwave \cite{Hardy:93} and $\mu$SR \cite{Sonier:94} measurements
on YBa$_2$Cu$_3$O$_y$ (YBCO) single crystals in the Meissner and vortex phases
established a limiting low-temperature linear $T$ dependence of $\lambda_{ab}$ 
that is consistent with $d$-wave pairing. More recently, 
non-$\mu$SR studies of YBCO in the Meissner phase have revealed that 
$T_c$ has a sublinear dependence on 
$1/\lambda_{ab}^2$.\cite{Pereg:04,Zuev:05,Broun:05}
On the other hand, the relation $T_c \! \propto \! \rho_s$ inferred from 
the Uemura plot is supported by a recent study of electric-field induced 
superfluid density modulations in a single underdoped ultra-thin film of 
La$_{2-x}$Sr$_x$CuO$_4$.\cite{Rufenacht:06}

The problem with assuming $\sigma \propto 1/\lambda_{ab}^2$ is that
there are additional inseparable contributions to $\sigma$ from electronic magnetic
moments and flux-line lattice (FLL) disorder, which may vary with doping. 
To circumvent this difficulty we have studied YBCO single crystals. 
In a single crystal the FLL contribution to the $\mu$SR line shape $n(B)$ 
is asymmetric and distinct from the other sources of field 
inhomogeneity.\cite{Sonier:00} Not only can the behavior of $\lambda_{ab}$ 
be isolated, but because the finite size of the vortex cores is apparent 
in a single-crystal measurement of $n(B)$, $\xi_{ab}$ can be simultaneously 
determined. While $\xi_{ab}$ may be accurately determined in conventional 
superconductors from measurements of the upper critical field $H_{c2}$,
in high-$T_c$ cuprates $H_{c2}$ is generally a very high magnetic field 
marking the transition from a vortex liquid to the normal phase.
Here we present $\mu$SR measurements that probe $\lambda_{ab}$ and 
$\xi_{ab}$ in the bulk of YBCO single crystals deep in the superconducting
state. The accuracy of our method was demonstrated in previous studies
of conventional superconductors,\cite{Sonier:04,Callaghan:05,Laulajainen:06} 
but is reinforced here through comparisons with the results from 
other techniques.     

\section{Experimental Details}

YBCO single crystals with purity of 99.995~\% were grown by a self-flux method 
in fabricated BaZrO$_3$ crucibles at the University of British Columbia.\cite{Liang:98}
An exception are $y = 6.60$ single crystals that were grown in 
yttria-stabilized-zirconia crucibles and characterized by a purity greater than
99.5~\%.\cite{Liang:92} Single crystals of Ca-doped YBCO were also
prepared in BaZrO$_3$ crucibles.
Typical sample sizes consisted of 3 to 5 single crystals from the same growth batch 
arranged in a mosaic to form a total $\hat{a}$-$\hat{b}$ surface area of 20-30~mm$^2$.
The thickness of the crystals are on the order of $\sim \! 0.1$~mm.
The superconducting transition temperatures of the single
crystals were measured using a SQUID magnetometer. 
Twin boundaries were removed from some of the higher-doped single crystals by applying 
pressure along the $\hat{a}$ or $\hat{b}$ directions at elevated temperature.
These basic sample characteristics are summarized in Table~\ref{Samples}. 

\begin{table}
\caption[Paramters]{Characteristics of the YBa$_2$Cu$_3$O$_y$
and (Y, Ca)Ba$_2$Cu$_3$O$_{6.98}$ single crystals. 
The hole concentration $p$ per CuO$_2$ layer
is determined from the dependence of $T_c$ on $p$ presented in 
Ref.~\cite{Liang:06} for similar YBCO single crystals.}  
\begin{center}
\begin{tabular}{cccc}
\hline \hline 
$y$ & $p$ & $T_c$ (K) & Detwinned \\
\hline
6.60 & 0.103 & 62.5 &   \\
6.57 & 0.110 & 59.0 &   \\
6.67 & 0.120 & 66.0 &   \\
6.75 & 0.132 & 74.6 & $\surd$ \\
6.80 & 0.140 & 84.5 & $\surd$ \\
6.95 & 0.172 & 93.2 & $\surd$  \\
                            \\
(Y,Ca)6.98 & 0.192 & 86.0 &  \\
\hline \hline
\end{tabular}
\label{Samples} 
\end{center}
\end{table} 

The $\mu$SR experiments were performed over a 3 year period on the M15 and 
M20B surface muon beam lines at TRIUMF, Vancouver, Canada. 
The $\mu$SR spectra were recorded
in a transverse-field (TF) geometry with the applied magnetic field 
{\bf H} perpendicular to the initial muon-spin polarization direction, 
and perpendicular to the $c$-axis of the single crystals. A 
TF-$\mu$SR spectrum comprised of 20 to 30 million muon decay events 
was taken at each temperature and magnetic field. 

\begin{figure}
\centerline{\epsfxsize=3.0in\epsfbox{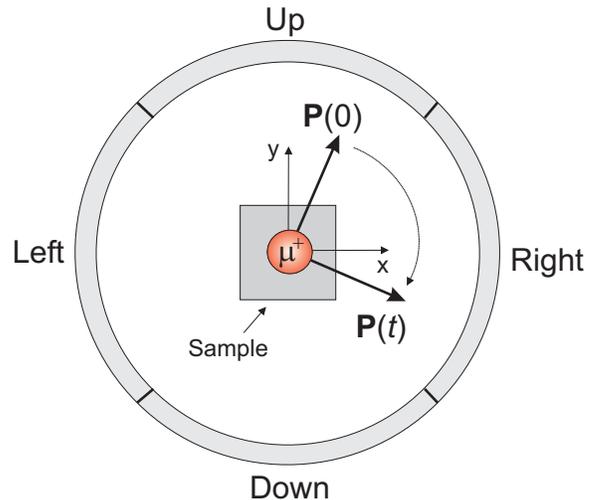}}
\caption{(Color online) Schematic of the positron counter arrangement
used in the present study. The muon beam axis and the applied 
magnetic field are perpendicular to the page ({\it i.e.}
parallel to the $z$-axis). The muon spin precesses in the $x$-$y$ plane 
about the local $z$-component of the magnetic field before
undergoing the decay $\mu^+ \! \rightarrow \! e^+ + \nu_e + \bar{\nu}_\mu$.
In our study, the muon spin polarization ${\bf P}(t)$
is formed from approximately 20 to 30 million muon decay events.
The decay events detected in the ``Left'' and ``Right'' positron 
counters form the $x$-component of ${\bf P}(t)$, while
decay events detected in the ``Up'' and ``Down'' positron
counters form the $y$-component.}
\label{fig1}
\end{figure}

In a transverse-field, the muon spin precesses in a plane perpendicular 
to the field direction (which we define here as the $z$-direction). 
The time evolution of the muon spin polarization
$P(t)$ is determined from the $\mu$SR ``asymmetry'' spectrum formed
from the muon decay events detected in opposing positron counters
\begin{equation}
A(t)  = \frac{N_1(t) - N_2(t)}{N_1(t) + N_2(t)} = a_0 P(t) \,  , 
\end{equation}
where $N_{\it i}(t)$ is the time histogram of the temporal dependence
of decay positron count rate in the {\it i}$^{th}$ detector, and 
$a_0$ is the asymmetry maximum. In our experiments four positron
counters were used to completely cover the 360$^\circ$ solid angle
in the $x$-$y$ plane (see Fig.~\ref{fig1}). The muon spin polarization 
function for the ``Left''-``Right'' pair of detectors is defined as 
\begin{equation}
P_x(t) = \int_0^{\infty} n(B) \cos(\gamma_\mu B t + \phi) dB \, ,
\label{eq:polarx}  
\end{equation}
and for the ``Up''-``Down'' pair as
\begin{equation}
P_y(t) = \int_0^{\infty} n(B) \cos(\gamma_\mu B t + \phi - \pi/2) dB \, ,
\label{eq:polary}  
\end{equation} 
where $\gamma_\mu$ is the muon gyromagnetic ratio,
$\phi$ is a phase constant, and
\begin{equation}
n(B^{\prime}) = \langle \delta [ B^{\prime} - B({\bf r})] \rangle \, ,
\label{distribution}
\end{equation}
is the probabilty of finding a local magnetic field $B$ in the $z$-direction 
at an arbitrary position {\bf r} in the $x$-$y$ plane. 

\section{Data Analysis Method}

The TF-$\mu$SR time spectra for each sample were fit assuming
the following analytical solution of the GL equations \cite{Yaouanc:97}
for the spatial field profile of the ideal FLL 
\begin{equation}
B({\bf r}) = B_0 \sum_{ {\bf G}}
\frac{e^{-i {\bf G} \cdot {\bf r}} \, \, F(G)}{\lambda_{ab}^2 G^2} \, ,
\label{eq:GLfield}
\end{equation}
where {\bf G} are the reciprocal lattice vectors, 
$B_0$ is the average internal magnetic field,
$F(G) = u K_1(u)$ is a cutoff function for the {\bf G} sum, 
$K_1(u)$ is a modified Bessel function, and 
$u = \sqrt{2} \xi_{ab} G$.
The cutoff function $F(G)$ depends on the spatial profile of the 
superconducting order parameter at the center of the vortex core.
Consequently, $\xi_{ab}$ is a measure of the {\it vortex core size}.
The FLL in all samples was assumed to be
hexagonal. Neutron scattering experiments on fully-doped
YBCO \cite{Brown:04} indicate that the FLL below 
$H \approx 40$ kOe is only slightly distorted from 
hexagonal symmetry due to $a/b$ anisotropy . 
We find that accounting for this small 
distortion changes the values of $\xi_{ab}$ and $\lambda_{ab}$ 
by less than 5~\%. Consequently, the FLL was
assumed to be hexagonal for all samples studied.

\begin{figure}
\centering
\centerline{\epsfxsize=5.0in\epsfbox{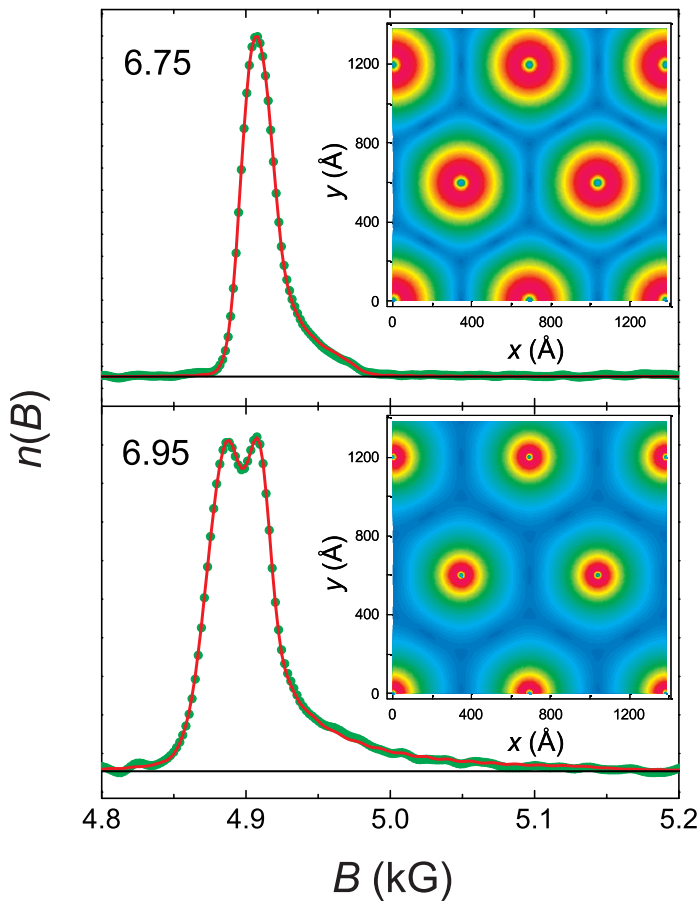}}
\caption{(Color online) $\mu$SR line shapes in single-crystal YBCO.
Fourier transforms of the TF-$\mu$SR signal from
$y = 6.75$ and $y = 6.95$ samples 
at $T \! = \! 2.5$~K and $H \! = \! 4.92$~kOe (green circles). 
The righthand peak for $y = 6.95$ is a background signal coming from 
muons stopping outside the sample (Note, the background and sample 
peaks nearly coincide in the $\mu$SR line shape for $y = 6.75$).
The red curves through the
data are the Fourier transforms of the fits in the time
domain. The contour plots show the corresponding spatial 
dependence of the supercurrent density 
$j(x, y) = |{\bf \nabla} \times {\bf B}(x, y)|$, 
providing a visual illustration of the change in core
size with hole doping.}
\label{fig2}
\end{figure} 

To avoid the difficulty of modelling the contribution of electronic
magnetic moments to the $\mu$SR line shape, we restricted our
study to YBCO crystals free of static or quasistatic spins. For the
applied fields considered in this study, this has been determined to 
be the case for oxygen content $y \! > \! 6.50$.\cite{Sonier:07} 
 
To properly account for disorder of the FLL, the dimensionality
of the vortices must be considered. Josephson plasma resonance measurements
on YBa$_2$Cu$_3$O$_{6.50}$ ortho-II single crystals grown by
Liang, Bonn and Hardy indicate that the vortices are 3D-like at low
temperaturesi,\cite{Dulic:01} while mutual inductance measurements 
on thin films by Zuev {\it et al} show that even severely
underdoped YBCO is quasi-2D only near $T_c$.\cite{Zuev:05} 
Since the focus in the
present study is on the variation of $\lambda_{ab}$ and $\xi_{ab}$
in higher doped samples at low temperatures, the vortices are
assumed to be rigid 3D lines of flux. This assumption is also 
consistent with the observation of highly asymmetric $\mu$SR line shapes 
for all of our samples at low $T$ (see Fig.~\ref{fig2}).
For rigid flux lines, random displacements of the vortices
from their positions in the 
ideal hexagonal FLL are accounted for by convoluting the theoretical
line shape $n(B)$ by a Gaussian distribution of fields.\cite{Brandt:88} 
A Gaussian function also describes the local distribution of dipolar
fields originating from static nuclear moments.\cite{Schenck:85}
Taking into account both sources of line broadening, the 
corresponding theoretical polarization functions are
\begin{eqnarray}
P_x(t) & = & e^{- \sigma_{\rm eff}^2t^2/2} 
\int_0^{\infty} n(B) \cos(\gamma_\mu B t + \phi) dB \, , \\
P_y(t) & = & e^{- \sigma_{\rm eff}^2t^2/2} 
\int_0^{\infty} n(B) \cos(\gamma_\mu B t + \phi - \pi/2) dB \, ,
\label{eq:Polartheory}  
\end{eqnarray}
where,
\begin{equation} 
\sigma_{\rm eff}^2 = \sigma_{\rm dip}^2 + \sigma_{\rm FLL}^2 \, ,
\end{equation}
is an effective depolarization rate due to nuclear dipole moments
($\sigma_{\rm dip}$) and FLL disorder ($\sigma_{\rm FLL}$). 
Values for $\sigma_{\rm dip}$ were obtained by
fitting the TF-$\mu$SR signal above $T_c$ to the 
theoretical polarization function
\begin{equation}
P(t) = e^{- \sigma_{\rm dip}^2t^2/2} \cos(\gamma_\mu B t + \phi) \, .
\label{eq:dipolar}  
\end{equation}
To account for the background signal from muons 
that did not stop in the sample, an additional term of the
form 
$(1-f) e^{- \sigma_{\rm B}^2t^2/2} \cos(\gamma_\mu Bt + \phi_{\rm B})$
was added to Eq.~(\ref{eq:Polartheory}) and to Eq.~(\ref{eq:dipolar}), 
where $f$ is the fraction of muons that stopped inside the sample. 
Values of $f$ for the different samples ranged from 0.8 to 0.9.

In the present work there are two marked 
improvements over the analysis done in our previous studies of
YBCO single crystals \cite{Sonier:97a,Sonier:99a,Sonier:99b}
that assumed the spatial field profile of Eq.~(\ref{eq:GLfield}):
(i) The earlier works used the asymptotic 
limit $K_1(u) = \sqrt{\pi/2u} \exp(-u)$
($u \gg 1$) for the Bessel function that appears in the cutoff 
$F({\bf G})$, whereas here $K_1(u)$ was evaluated numerically.
(ii) The second improvement is that due to increased computer speed,
$B({\bf r})$ was calculated at 15,132 equally-spaced locations 
in the rhombic unit cell of the hexagonal FLL, compared to
5,628 locations in previous works.
Further increasing the number of real-space points sampled
in the FLL unit cell did not result in appreciable changes
in the fitted parameters.
We note that both improvements in our data analysis method 
influence the absolute values of $\lambda_{ab}$ and $\xi_{ab}$,
but the temperature and magnetic field dependences of these 
parameters remain qualitatively similar to
that determined in our previous studies.

\section{Results for $\lambda_{ab}$}

\subsection{Temperature dependence}

\begin{figure}
\centerline{\epsfxsize=4.4in\epsfbox{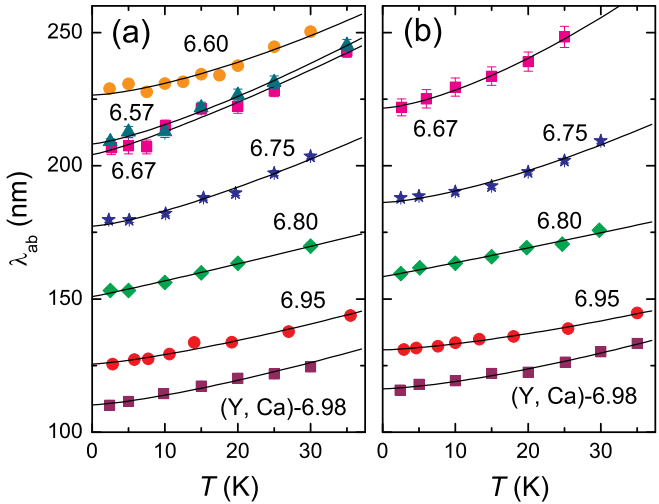}}
\caption{(Color online) 
Temperature dependence of $\lambda_{ab}$ at (a) $H \! = \! 5$~kOe
and (b) $H \! = \! 15$~kOe. The solid curves 
are fits to $\lambda_{ab}(T, H) \! = \! \lambda_{ab}(0, H) + \alpha T^{n}$, 
where $\alpha$ and $n$ are field-dependent coefficients.}
\label{fig3}
\end{figure}

\begin{figure}
\centerline{\epsfxsize=4.0in\epsfbox{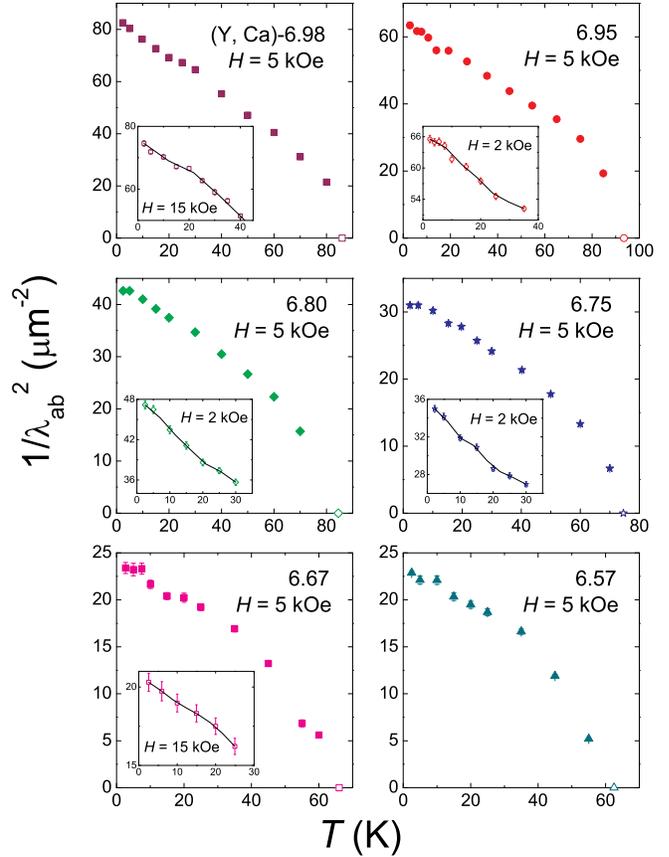}}
\caption{(Color online) Temperature dependence of $1/\lambda_{ab}^2$ at 
$H \! = \! 5$~kOe. The insets show additional low temperature
measurements at $H \! = \! 2$~kOe or $H \! = \! 15$~kOe.
The solid curves through the data points are merely guides
for the eye.}
\label{fig4}
\end{figure}

Figure~\ref{fig3} shows the temperature dependence of
$\lambda_{ab}$ at low $T$ determined at two different values of the
applied magnetic field. The solid curves 
are fits to $\lambda_{ab}(T, H) \! = \! \lambda_{ab}(0, H) + \alpha T^{n}$, 
where $\alpha$ and $n$ are field-dependent coefficients.
The dependence of $1/\lambda_{ab}^2$ on $T$ is
shown in Fig.~\ref{fig4} for selected values of the applied field.
An inflection point at $T \! \approx \! 20$~K is visible in some
of the lower field data. This feature was also apparent 
in our previous measurements 
of YBa$_2$Cu$_3$O$_{6.95}$.\cite{Sonier:99b}
Harshman {\it et al} have argued that the inflection
point is caused by thermal depinning of vortices,\cite{Harshman:04} 
although an invalid treatment of the data was used to support 
this assertion.\cite{Sonier:05}
Recently, Khasanov {\it et al} have ruled out
depinning as the source of a similar inflection point in the temperature 
dependence of $1/\lambda_{ab}^2$ measured in La$_{1.83}$Sr$_{0.17}$CuO$_4$
by TF-$\mu$SR.\cite{Khasanov:07} 
Instead they attribute this
feature to the occurrence of both a large $d$-wave and a 
small $s$-wave superconducting gap. As we will show later,
the anomalous magnetic field dependence of the vortex core
size in YBCO can be explained by an induced superconducting
energy gap in the CuO chains that run along the $b$ direction. 
Theoretical calculations by
Atkinson and Carbotte \cite{Atkinson:95} for a $d$-wave superconductor 
with proximity-induced superconductivity in the CuO chains, show 
that $1/\lambda_{ab}^2(T)$ exhibits an inflection point 
caused by an upturn of $1/\lambda_{b}^2(T)$ at low $T$ (where
$\lambda_b$ is the penetration depth in the $b$ direction).                         

\begin{figure}
\centerline{\epsfxsize=4.0in\epsfbox{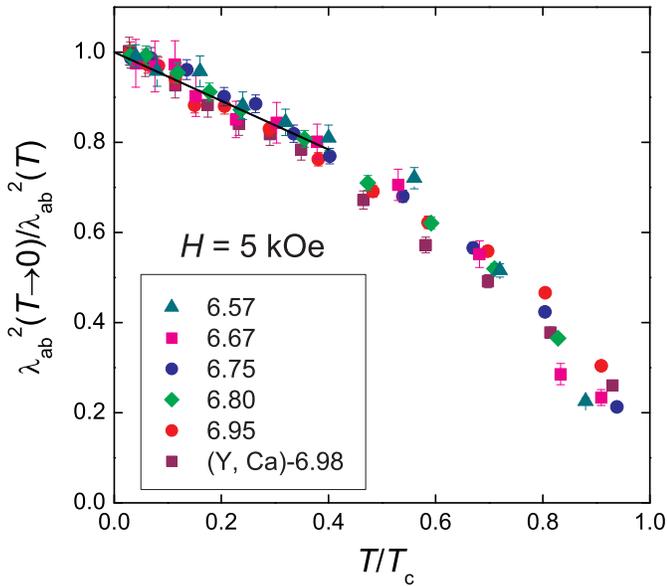}}
\caption{(Color online) Temperature dependence of $1/\lambda_{ab}^2$ at $H \! = \! 5$~kOe, 
normalized to $1/\lambda_{ab}^2(T \! \rightarrow \! 0)$ and plotted as a 
function of reduced temperature $T/T_c$. The solid line through the data
is a fit given by $\lambda_{ab}^2(T \rightarrow 0)/\lambda_{ab}^2(T) = 1 - 0.54 T/T_c$.}
\label{fig5}
\end{figure}

In Fig.~\ref{fig5}
it is shown that $\lambda_{ab}^2(T \! \rightarrow \! 0)/ \lambda_{ab}^2(T)$ 
exhibits a near universal linear temperature dependence at low $T$. 
We attribute deviations
from universal behavior near $T_c$ to softening of the FLL,
which narrows the $\mu$SR line shape and enhances the fitted
value of $\lambda_{ab}$.
The universal scaling implies that 
$\lambda_{ab}^2(T \! \rightarrow \! 0) T_c v_{\rm F} Z^{e2}/v_{\Delta}$ is
a constant,\cite{Ioffe:02a} 
where $v_{\rm F}$ is the Fermi velocity, $v_{\Delta}$ is a velocity
corresponding to the slope of the gap at the nodes, and $Z^e$ is a 
charge renormalization parameterizing the coupling 
of the quasiparticles to phase fluctuations.
Using values of $v_{\Delta}$ from thermal conductivity 
measurements,\cite{Sutherland:03} we find that $Z^e$ is bascially 
doping independent. 

\subsection{Magnetic field dependence}

\begin{figure}
\centerline{\epsfxsize=3.5in\epsfbox{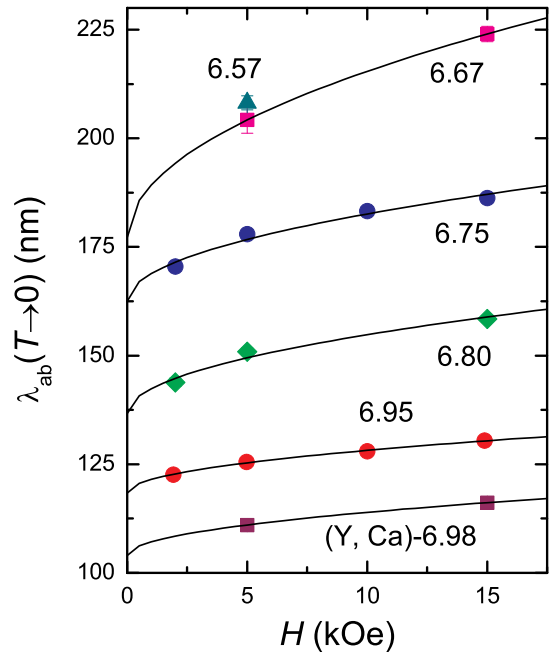}}
\caption{(Color online)
Magnetic field dependence of the extrapolated zero-temperature value of 
$\lambda_{ab}$. The solid curves are fits to 
$\lambda_{ab}(0, H) \! = \! \lambda_{ab}(0, 0) 
+ \beta \sqrt{H}$, where the coefficient $\beta$ decreases with increasing hole doping.}
\label{fig6}
\end{figure}

Figure~\ref{fig6} shows the magnetic field dependences of
$\lambda_{ab}$. Here we stress that the observed behaviors do not
imply that the magnetic penetration depth or superfluid density
depend on field in this way. The sublinear dependence of $\lambda_{ab}$ 
on $H$ is primarily due to the failure of Eq.~(\ref{eq:GLfield})
to account for all field-dependent contributions to the 
internal magnetic field distribution.
In Refs.~\cite{Sonier:99b,Amin:98} the strong field dependence of 
$\lambda_{ab}$ in YBa$_2$Cu$_3$O$_{6.95}$ determined by $\mu$SR
was explained by the high anisotropy of the $d$-wave superconducting 
energy gap not accounted for in Eq.~(\ref{eq:GLfield}).
A nonlocal supercurrent response to the
applied field in the vicinity of the vortex cores stemming 
from the divergence of the coherence length at the gap nodes 
modifies the spatial distribution of field. With increasing $H$,
the increased overlap of the regions around the vortex cores
reduces the width of the $\mu$SR line shape.
The gap anisotropy also results in a nonlinear supercurrent
response to the applied field, resulting from a quasiclassical
`Doppler shift' of the quasiparticle energy spectrum by the
flow of superfluid around a vortex.\cite{Volovik:93} 
When the Doppler shift exceeds the energy gap, Cooper pairs are 
broken, and $\lambda_{ab}$ increases.

These effects are not restricted to $d$-wave superconductors.
Sizeable nonlinear and/or nonlocal effects can also occur in
$s$-wave superconductors with a smaller energy gap on one of the Fermi 
sheets and/or a highly anisotropic Fermi surface.
Moreover, these anisotropies result in a rapid delocalization
of quasiparticle core states with increasing $H$ that modify
$n(B)$. Indeed, strong field dependences of $\lambda_{ab}$ from
Eq.~(\ref{eq:GLfield}) have been observed in the multi-band 
superconductor NbSe$_2$,\cite{Callaghan:05} and the
marginal type-II superconductor V.\cite{Laulajainen:06}
It has been experimentally established 
for a variety of materials,\cite{Callaghan:05} 
including YBCO,\cite{Pereg:04} that the $H \rightarrow 0$ 
extrapolated value of $\lambda_{ab}$ agrees 
with the magnetic penetration depth measured by other techniques 
in the Meissner phase.
Consequently, we stress that only $\lambda_{ab}(H \rightarrow 0)$ can be
considered a ``true'' measure of the magnetic penetration depth.      

\subsection{Hole doping dependence}

\begin{figure}
\centerline{\epsfxsize=4.0in\epsfbox{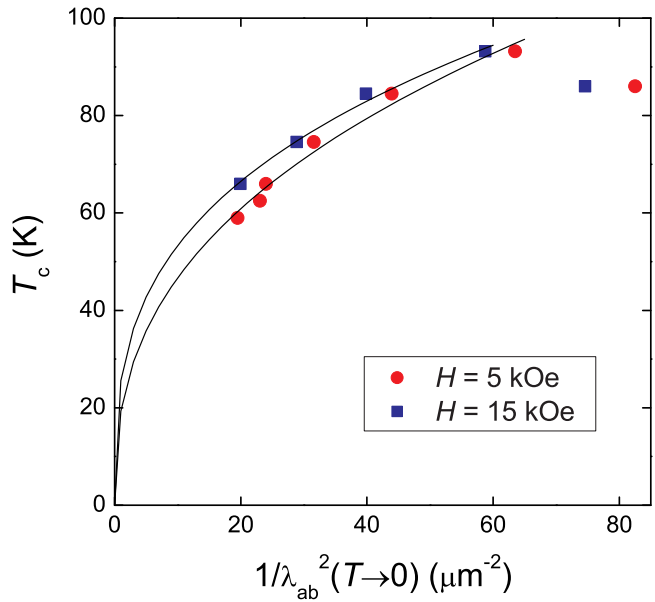}}
\caption{(Color online) Dependence of $T_c$ on 
$1/\lambda_{ab}^2(T \! \rightarrow \! 0)$.
The two data points on the far right are 
for the overdoped sample (Y,Ca)Ba$_2$Cu$_3$O$_{6.98}$.  
The solid curves are fits to the data for YBa$_2$Cu$_3$O$_y$, 
yielding $T_c \! = \! (19.3 \! \pm \! 1.7$~K$\cdot \mu$m$^2) 
[1/\lambda_{ab}^2]^{0.38 \pm 0.02}$ and 
$T_c \! = \! (25.5 \! \pm \! 1.9$~K$\cdot \mu$m$^2) 
[1/\lambda_{ab}^2]^{0.32 \pm 0.02}$ at $H \! = \! 5$~kOe and 
$H \! = \! 15$~kOe, respectively.}
\label{fig7} 
\end{figure} 

In Fig.~\ref{fig7} we show $T_c$ as a function of 
$1/ \lambda_{ab}^2(T \rightarrow 0)$ at two different fields. The more inclusive
data set at $H = 5$~kOe is described by $T_c \! \propto \! (1/ \lambda_{ab}^2)^{0.38}$,
which deviates substantially from the linear scaling in the Uemura plot.
The power 0.38 is surprisingly close to 0.43 determined by 
Zuev {\it et al} in a Meissner phase study of severely 
underdoped YBCO thin films.\cite{Zuev:05} 
It is surprising because these thin films
have a superfluid density that is significantly
lower than in single crystals.\cite{Broun:05} 
It is known from microwave studies of YBCO that the doping 
dependences of $\lambda_{a}$ and $\lambda_{b}$ are not the same,\cite{Pereg:04}
due to the conductivity of the CuO chains.\cite{Atkinson:95}
Thus there is no reason to expect the power $\sim 0.4$ 
to be universal for the cuprates. 
While a sublinear dependence of $T_c$ on $1/ \lambda_{ab}^2$ has 
also been inferred from more recent $\mu$SR measurements of 
the muon depolarization rate $\sigma$ 
in other high-$T_c$ superconductors,\cite{Tallon:03} the contributions of 
magnetism and FLL disorder to the $\mu$SR line shape were not factored out.

\begin{figure}
\centerline{\epsfxsize=3.8in\epsfbox{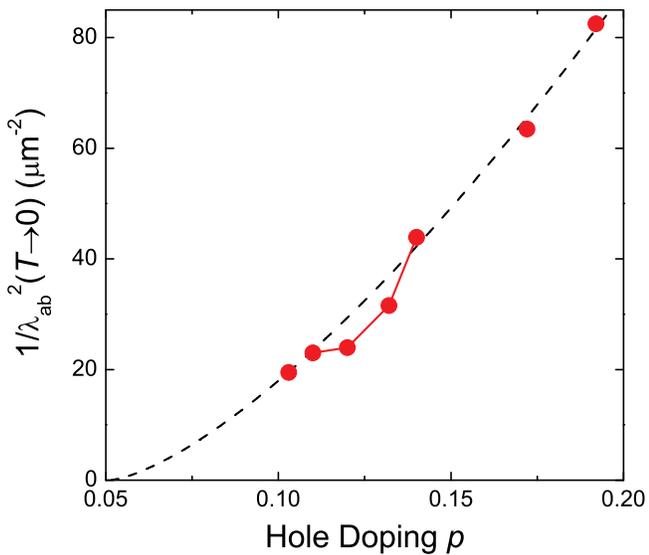}}
\caption{(Color online) Dependence of $1/\lambda_{ab}^2(T \! \rightarrow \! 0)$ 
on hole doping concentration $p$ at $H = 5$ kOe.
The dashed curve is the function $1/ \lambda_{ab}^2 \propto (p - 0.05)^{1.42}$,
where $p \! = \! 0.5$ is the critical hole doping 
concentration for the onset of superconductivity.}
\label{fig8} 
\end{figure} 

Figure~\ref{fig8} shows $1/ \lambda_{ab}^2$ as a function of hole
doping, where the values of $p$ are determined from the dependence
of $T_c$ on $p$ presented in Ref.~\cite{Liang:06} for similar YBCO single 
crystals. The behavior is consistent with 
other studies on cuprates indicating that the maximum value 
of $1/ \lambda_{ab}^2$ is not reached
before $p \approx 0.19$.\cite{Bernhard:01,Panagopoulos:02}    
Our data are in the range $0.103 \leq p \leq 0.192$ 
and are described by $1/ \lambda_{ab}^2 \propto (p - 0.05)^{1.42}$, except 
near $p = 0.125 = 1/8$. 
 
\begin{figure}
\centerline{\epsfxsize=4.0in\epsfbox{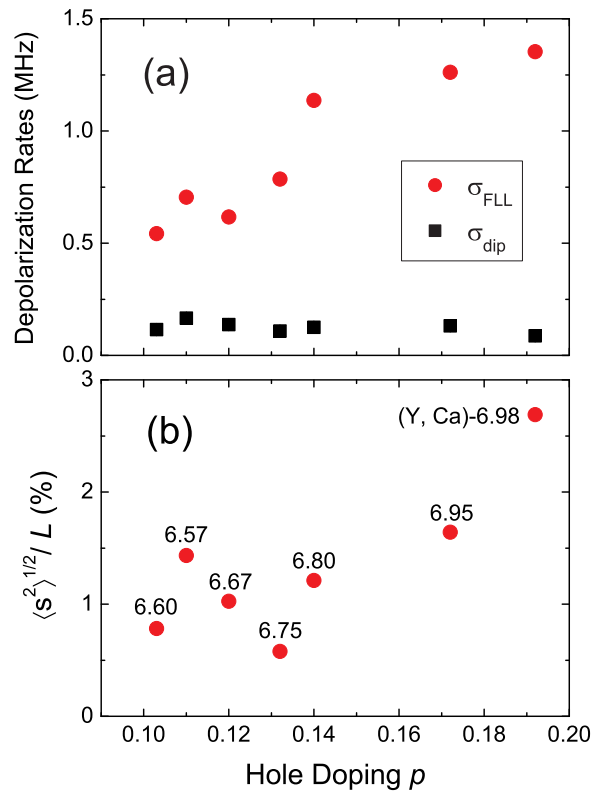}}
\caption{(Color online) (a) Hole doping dependences of the Gaussian 
depolarization rates $\sigma_{\rm FLL}$ and $\sigma_{\rm dip}$ 
at $T \rightarrow 0$ K and $H = 5$ kOe. 
(b) Hole doping dependence of the root-mean-square 
displacement $\langle s^2 \rangle^{1/2}$ of the vortices from their 
positions in the perfect hexagonal FLL at $T \rightarrow 0$ K and 
$H = 5$ kOe plotted as a percentage of the intervortex spacing $L$.}
\label{fig9} 
\end{figure}

The hole doping dependences of $\sigma_{\rm dip}$ and $\sigma_{\rm FLL}$
at $H = 5$ kOe are shown in Fig.~\ref{fig9}(a). While $\sigma_{\rm dip}$ is
independent of $p$, $\sigma_{\rm FLL}$ basically tracks $1/\lambda_{ab}^2$.
Using the fitted values of $\sigma_{\rm FLL}$ 
and $\lambda_{ab}$ we have calculated the hole doping dependence of the
root-mean-square displacement $\langle s^2 \rangle^{1/2}$ of the 
vortices from there positions in the perfect hexagonal FLL. 
As shown in Fig.~\ref{fig9}(b), the degree of FLL disorder is small and 
as expected highest in the Ca-doped sample.  

\section{Results for $\xi_{ab}$}

\subsection{Magnetic field dependence}

\begin{figure}
\centerline{\epsfxsize=3.8in\epsfbox{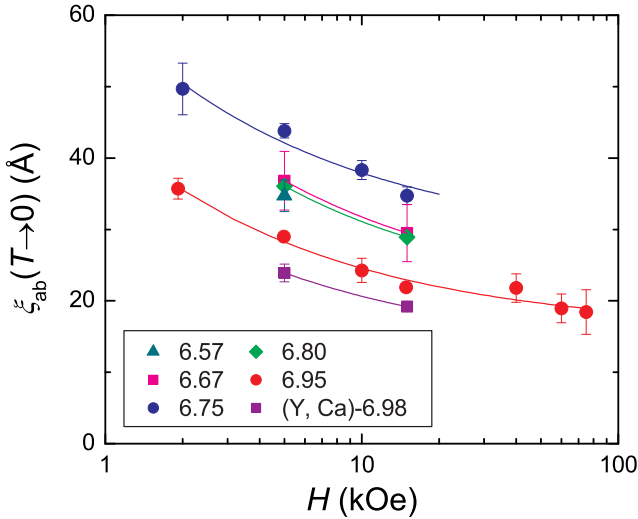}}
\caption{Magnetic field dependence of the extrapolated 
zero-temperature value of $\xi_{ab}$. The solid curves are fits to 
$\xi_{ab}(0,H) = a + b/\sqrt{H}$, where the 
coefficients $a$ and $b$ depend on the hole doping concentration.}
\label{fig10}
\end{figure}

The field dependences of $\xi_{ab}$ are shown in Fig.~\ref{fig10}.
The increase in $\xi_{ab}$ at low field, which corresponds to
an expansion of the vortex cores, was previously reported
for YBa$_2$Cu$_3$O$_{6.60}$ (Ref.~\cite{Sonier:97a}) and
YBa$_2$Cu$_3$O$_{6.95}$ (Refs.~\cite{Sonier:99a,Sonier:99b}).
In all samples we find that $\xi_{ab}$ scales as $1/\sqrt{H}$, which
is proportional to the intervortex spacing. In $s$-wave
superconductors the field dependence of $\xi_{ab}$ has been 
shown to originate from the delocalization of bound quasiparticle 
core states.\cite{Callaghan:05,Sonier:04} This is because a
change in the spatial dependence of the pair potential 
accompanies the change in electronic structure of
the vortex cores. In YBCO the low-energy quasiparticle core states 
should be extended along the nodal directions of the $d$-wave gap 
function.\cite{Ichioka:96} This allows for a large transfer of 
low-energy quasiparticles between vortices at low field, which is 
further enhanced by an increase in vortex density. Hence the vortex 
core size is predicted to shrink with 
increasing field.\cite{Ichioka:99a,Ichioka:99b} However,
the field dependence of $\xi_{ab}$ in YBCO is considerably
stronger than predicted for a pure $d$-wave superconductor.
Consequently, we consider an alternative explanation.

\begin{figure}
\centerline{\epsfxsize=3.8in\epsfbox{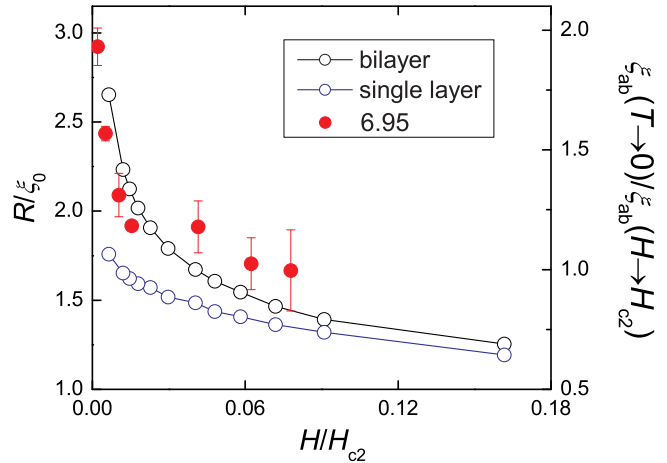}}
\caption{Calculated field-dependent core size $R$ relative to
the BCS coherence length $\xi_0$ for single-layer and bilayer models
(see Appendix).
Also shown is the field dependence of $\xi_{ab}(T \rightarrow 0)$
in the $y = 6.95$ sample normalized to the extrapolated value
$\xi_{ab}(H \rightarrow H_{c2}) = 18.5$~\AA~ and plotted as a 
function of $H/H_{c2}$, where 
$H_{c2} = \Phi_0/2 \pi \xi_{ab}^2(H \rightarrow H_{c2})$.}
\label{fig11}
\end{figure}

In Fig.~\ref{fig11} we show that the field-dependence of $\xi_{ab}$, in
particular the upturn at low field, can be explained by the presence
of the CuO chains. The calculations of the core size $R$ are based on a
semiclassical Doppler-shift approximation (see Appendix) for either a
single-layer model representing a superconducting CuO$_2$ plane or a
proximity-coupled model representing a CuO$_2$-CuO bilayer.  In the
bilayer model there are two distinct energy scales for pair breaking:
the energy gap associated with Cooper pairs in the CuO$_2$ planes, and
a smaller proximity-induced gap associated with the chains.  
It is the latter scale which is responsible for the expansion of
the vortex cores at low field.

\begin{figure}
\centerline{\epsfxsize=3.8in\epsfbox{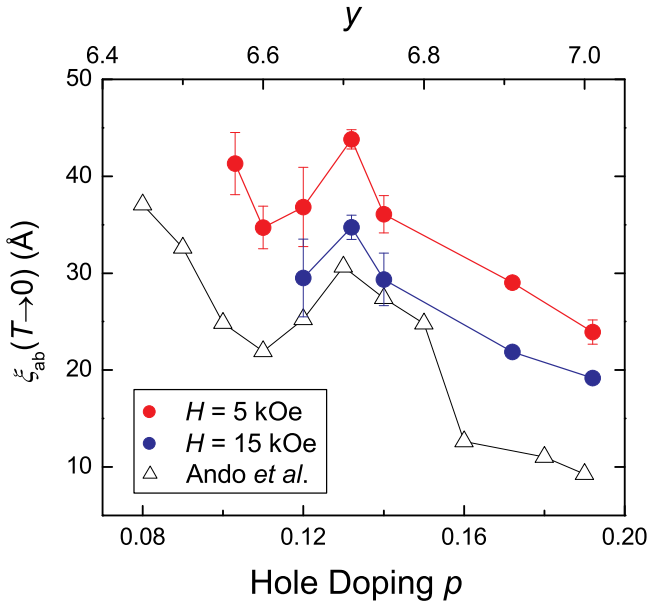}}
\caption{Hole doping dependence of $\xi_{ab}(T \rightarrow 0)$
at $H = 5$ kOe and $H = 15$ kOe. Also shown is data for
$\xi_{ab}$ from Ref.~\cite{Ando:02} plotted as a function of
$y$, rather than $p$.}
\label{fig12}
\end{figure}

\begin{figure}
\centerline{\epsfxsize=4.0in\epsfbox{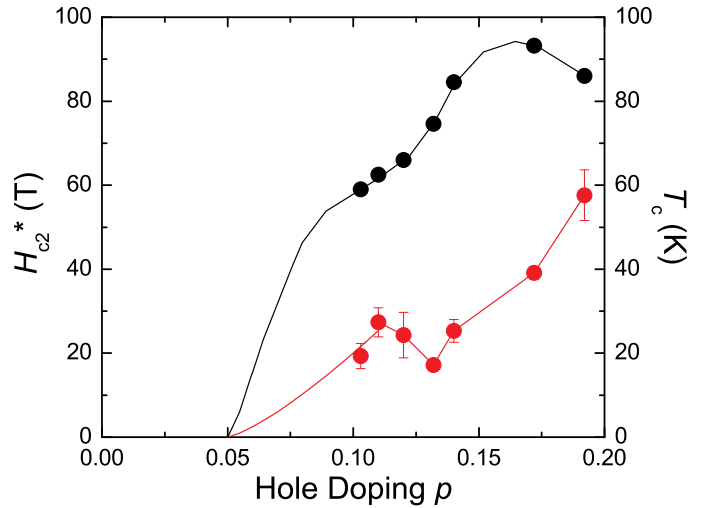}}
\caption{Plot of $H^*_{c2} = \Phi_0/2 \pi \xi_{ab}^2(T \rightarrow 0)$
as a function of hole doping (red circles), using the values of 
$\xi_{ab}(T \rightarrow 0)$ at $H = 5$ kOe. Also shown are the
values of $T_c$ (black circles). The black curve is the
relation between $T_c$ and $p$ from Ref.~\cite{Liang:06}.}
\label{fig13}  
\end{figure} 

\subsection{Hole doping dependence}

Figure~\ref{fig12} shows the hole doping dependences of
$\xi_{ab}(T \! \rightarrow \! 0)$ at $H \! = \! 5$~kOe 
and $H \! = \! 15$~kOe. 
Qualitatively, the doping dependence of $\xi_{ab}$ is similar
to that reported by Ando {\it et al} from magnetoconductance 
measurements on detwinned YBCO single crystals.\cite{Ando:02}
This result is also shown in Fig.~\ref{fig12}, but plotted as
$\xi_{ab}$ versus $y$. Note that our data must be plotted as
$\xi_{ab}$ versus $p$, because the hole doping concentration
of our $y = 6.60$ single crystals (grown in a different 
kind of crucible than the other samples), is smaller than 
that of $y = 6.57$ (see Table~\ref{Samples}).  
The general trend of all data sets is an increase of $\xi_{ab}$ 
with decreasing $p$. Such behavior has also been observed in the
underdoped regime of La$_{2-x}$Sr$_x$CuO$_4$,\cite{Wen:03,Kadono:04,Wang:06} 
and Bi$_2$Sr$_2$CuO$_{6+\delta}$.\cite{Bouquet:06} 
With increasing magnetic field, our values for
$\xi_{ab}$ approach those determined by Ando {\it et al}.
Note that based on our proximity-induced model for 
the field dependence of $\xi_{ab}$ (see Appendix), 
it is the high-field values of $\xi_{ab}$ 
that reflect the intrinsic superconductivity of the CuO$_2$ planes.  

Since the doping dependences of $\xi_{ab}$ at $H = 5$ kOe and 
$H = 15$ kOe in Fig.~\ref{fig12} are similar, we expect the
hole doping dependence of $H^*_{c2} \equiv \Phi_0/2 \pi \xi_{ab}^2$ 
to qualitatively resemble that of the upper critical field $H_{c2}$.
Figure~\ref{fig13} shows the hole doping dependence of $H^*_{c2}$
calculated from the values of $\xi_{ab}$ at $H = 5$ kOe. 
Consistent with the data of Ando {\it et al}, $H_{c2}$ decreases 
with decreasing $p$ in the underdoped regime of YBCO, and displays 
a dip near 1/8 hole doping. 

\section{Summary and Conclusions}

We have simultaneously determined the hole doping dependences
of the magnetic penetration depth and the GL coherence length
in the underdoped regime of YBa$_2$Cu$_3$O$_y$. This was achieved
by fitting $\mu$SR measurements in the vortex state to an analytical
solution of the GL equations for the internal magnetic field 
distribution. In this type of analysis the magnetic penetration depth is
strictly defined as the $H \! \rightarrow \! 0$ extrapolated value
of the fitted parameter $\lambda_{ab}$. The accuracy of this definition 
was established in previous studies of conventional 
superconductors.\cite{Callaghan:05,Laulajainen:06} Here we
have presented measurements showing a refinement of the Uemura
plot for YBa$_2$Cu$_3$O$_y$, where $T_c$ is plotted as a function of
the isolated quantity $1/\lambda_{ab}^2$, rather than the
the muon depolarization rate $\sigma$. We find that
$T_c$ exhibits a strong sublinear dependence on $1/\lambda_{ab}^2$, 
suggesting that $T_c$ is not directly proportional to the 
superfluid density $\rho_s$. This result supports the same conclusion
reached in several recent Meissner phase studies of YBCO.

We have reported here a reduction of $1/\lambda_{ab}^2$ 
near 1/8 hole doping concentration. Suppression of $1/ \lambda_{ab}^2$ or 
the muon depolarization rate $\sigma$ near 1/8 hole doping has 
only previously been observed in cuprates where $p$ is controlled by 
cation substitution,\cite{Bernhard:01,Panagopoulos:02} and was
believed to indicate a tendency toward static stripe formation.
Static stripes over the doping range investigated here were recently 
ruled out by inelastic neutron scattering experiments on YBCO.\cite{Hinkov:04}
We ourselves find no evidence for static spins in  
zero-field $\mu$SR or TF-$\mu$SR measurements on our 
samples.\cite{Sonier:07}  
However, the suppression of superconductivity near $p = 1/8$ 
could be caused by fluctuating stripes, recently 
argued to be relevant in YBCO and other cuprates.\cite{Vojta:06}
Experimental evidence for dynamic stripes in YBCO includes the
detection of low-energy one-dimensional incommensurate modulations
in YBa$_2$Cu$_3$O$_{6.50}$ by inelastic neutron scattering.\cite{Stock:04}  

Further evidence for suppression of 
superconductivity near $p \! = \! 1/8$ is found in the hole 
doping dependence of $\xi_{ab}$. 
In our measurements $\xi_{ab}$ is a parameter that characterizes 
the size of the vortex cores. While it mimics the behavior of the
GL coherence length, $\xi_{ab}$ is large at low field due to
the contribution of the CuO chains to the spatial dependence of
the superconducting order parameter. Enhancement of the
GL coherence length or vortex core size near 1/8 hole doping, 
has also been observed in La$_{2-x}$Sr$_x$CuO$_4$.\cite{Wen:03}   
Calculations by Mierzejewski and M\'{a}ska show that
static or quasistatic stripes actually intensify $H_{c2}$ by
reducing diamagnetic pair breaking,\cite{Mierz:02} and
hence cannot explain the growth of $\xi_{ab}$ near $p \! = \! 1/8$. 
On the other hand, Kadono {\it et al} have shown that an
expansion of the vortex cores with decreasing hole doping 
can result from a strengthening of antiferromagnetic correlations 
competing with superconductivity.\cite{Kadono:04}
Thus, dynamic stripes are a viable explanation
for the increased size of the vortex cores near 1/8 hole doping.
  
\section*{Acknowledgements}  
We gratefully acknowledge D.~Broun, I.~Vekhter and A.~J.~Millis
for helpful and informative discussions. We also thank Y.~Ando for allowing us 
to reproduce his data here. This work was supported by the Canadian 
Institute for Advanced Research 
and the Natural Sciences and Engineering Research Council (NSERC) of Canada.\\

\section*{Appendix: Semiclassical Calculation of the Vortex Core Size}
Calculations of the vortex core size are based on a generalisation of
the so-called ``doppler-shift'' approximation for the vortex
structure \cite{Vekhter:01,Knapp:01} to the case of YBCO, which is a multiband
superconductor.  
In the case of YBCO, there is strong evidence that
both the two-dimensional CuO$_2$ planes and one-dimensional CuO chains
superconduct.  
Furthermore, it is likely that the chains are intrinsically normal, but are
driven superconducting by the proximity effect.
Proximity models for YBCO have been extensively described
elsewhere \cite{Atkinson:95,Atkinson:99}.
The essential idea is that the
superconductivity originates from a pairing interaction which is
confined to the two-dimensional CuO$_2$ planes, and that the mixing
of chain and plane wavefunctions induces superconductivity in the
one-dimensional chain layers.  

We adopt a simplified bilayer model consisting of a single plane and a
single chain, with one Wannier orbital retained per unit cell for each
layer.  For comparison purposes, calculations are also performed for a
single-layer model of an isolated superconducting plane.  The
Bogoliubov-deGennes Hamiltonian for the bilayer is
\begin{equation}
\hat H = \sum_{ij} \hat \Psi_i^\dagger
\left [ \begin{array}{cccc}
{\tilde t}_{1,ij} & \Delta_{ij} & t_\perp\delta_{i,j} & 0 \\
\Delta^\dagger_{ij} & -{\tilde t}_{1,ij}{}^\ast & 0 & - t_\perp \delta_{i,j} \\
t_\perp \delta_{i,j} & 0 & {\tilde t}_{2,ij} & 0 \\
0& -t_\perp \delta_{i,j} & 0 & - {\tilde t}_{2,ij} {}^\ast 
\end{array} \right ] \hat \Psi_j
\label{eq:ham}
\end{equation}
where $\hat \Psi^\dagger_i = [\psi^\dagger_{1i\uparrow}
\psi_{1i\downarrow} \psi_{2i\uparrow}^\dagger \psi_{2i\downarrow} ]$
and $\psi^\dagger_{ni\uparrow}$ ($\psi_{ni\downarrow}$) are creation
operators for quasiparticles (quasiholes) at lattice site $i$ in layer
$n$.  Here, we take $n=1$ for the plane layer and $n=2$ for the chain
layer.  The parameters $\tilde t_{n,ij}$ and $t_\perp \delta_{i,j}$ are the
single-electron hopping matrix elments between sites $i$ and $j$
within and between layers respectively, while $\Delta_{ij}$ is the
superconducting order parameter along bonds connecting nearest
neighbour sites $i$ and $j$.  From the form of Eq.~(\ref{eq:ham}), it
is apparent that $\Delta_{ij}$ only couples quasiparticles
belonging to the plane layer.  The single-layer Hamiltonian is obtained by
setting $t_\perp =0$.

A magnetic field $H$ applied perpendicular to the layers induces
circulating currents in the superfluid.  
The superfluid velocity is
given by ${\bf v_s} = {\bf M}^{-1} \cdot {\bf p_s}$ where ${\bf M}$ is
the effective mass tensor, ${\bf p_s}({\bf r}) = (2e\hbar/c){\bf A}({\bf
r}) + \hbar\nabla \phi({\bf r})$ is the superfluid momentum, ${\bf A}({\bf
r})$ is the magnetic vector potential and $\phi({\bf r})$ is the local
phase of the order parameter.  In the limit that $\phi({\bf r})$ and
${\bf A}({\bf r})$ are slowly varying functions, one can treat the
superflow as uniform in the neighbourhood of ${\bf r}$.  Then, one can
make a local gauge transformation \cite{Vekhter:01,Knapp:01} such
that the phase is removed from the order parameter and appears instead
in the hopping matrix elements ${\tilde t}_{n,ij}$:
\begin{eqnarray} 
{\tilde t}_{n,ij} &=& t_{n,ij} e^{-i{\bf p_s}({\bf r})\cdot 
({\bf r}_i-{\bf r}_j)/2\hbar}
\label{eq:nonlinear} \\
&\approx& t_{n,ij} +  \frac{1}{2} {\bf v}_{ij} \cdot {\bf p_s}({\bf r})
\label{eq:linear}
\end{eqnarray}
where $t_{n,ij}$ are the hopping matrix elements in zero-field and
 ${\bf v}_{ij} = -i t_{n,ij} ({\bf r}_{i}-{\bf r}_j)/\hbar$ are the
 matrix elements of the zero-field quasiparticle velocity.  Equation
 (\ref{eq:linear}) follows from Eq.~(\ref{eq:nonlinear}) in the limit
 that ${\bf p_s}$ is small.  Then, the order parameter takes on the
 simple $d$-wave form $\Delta_{ij} = \frac{1}{2}\Delta (-1)^{y_i-y_j}$
 which, in reciprocal space, corresponds to $\Delta_{\bf k} =
 \Delta[\cos(k_xa) - \cos(k_ya)]$, where $a$ is the lattice constant.
 The local gauge transformation leads to a doppler-shifted spectrum
 and is exact in the limit of slowly varying superfluid velocity.
 This procedure has been shown, in many circumstances, to provide a
 reasonable description of the vortex lattice \cite{Vekhter:01,Knapp:01}.

We take band structures which are appropriate for YBCO and adopt
\begin{equation}
t_{1,ij} = \left \{ \begin{array}{ll}
t_0 & i=j\\
t_1 & i,j \mbox{ are nearest neighbours} \\
t_2 & i,j \mbox{ are next-nearest neighbours}
\end{array}
\right. ,
\end{equation}
and
\begin{equation}
t_{2,ij} = \left \{ \begin{array}{ll}
t_3 & i=j\\
t_4 & i,j \mbox{ are nearest neighbours along }{\bf \hat y} 
\end{array}
\right. .
\end{equation}
For this work, we measure energies in units of $|t_1|$ and take
$\{t_0,\ldots,t_4\} = \{1,-1,0.45,2,-4\}$. In reciprocal space, the
dispersions of the isolated plane and chain layers are then
$\epsilon_{1{\bf k}} =
t_0+2t_1[\cos(k_xa)+\cos(k_ya)]+4t_2\cos(k_xa)\cos(k_ya)$ and
$\epsilon_{2{\bf k}} = t_3 + 2t_4\cos(k_ya)$, respectively. The
chain-plane hopping matrix element $t_\perp$ is not well known in YBCO
and is taken to be $t_\perp = 0.75$.

\begin{widetext}
 For a slowly varying ${\bf p_s}({\bf r})$ we can locally Fourier
transform the Hamiltonian in the neighbourhood of ${\bf r}$ to give
\begin{equation}
\hat H({\bf r}) = \sum_{k} \hat \Psi_k^\dagger
\left [ \begin{array}{cccc}
\epsilon_{1k} +\frac 12 {\bf v}_{1k}\cdot {\bf p_s}({\bf r}) & \Delta_k({\bf r}) & t_\perp & 0 \\
\Delta_k({\bf r}) & -\epsilon_{1k} +\frac 12 {\bf v}_{1k}\cdot {\bf p_s}({\bf r}) & 0 & - t_\perp \\
t_\perp  & 0 & \epsilon_{2k} +\frac 12 {\bf v}_{2k}\cdot {\bf p_s}({\bf r}) & 0 \\
0& -t_\perp  & 0 & - \epsilon_{2k} +\frac 12 {\bf v}_{2k}\cdot {\bf p_s}({\bf r}) 
\end{array} \right ] \hat \Psi_k
\label{eq:ham2}
\end{equation}
\end{widetext}
where ${\bf v}_{n{\bf k}} = \hbar^{-1}\partial \epsilon_{n{\bf
k}}/\partial{\bf k}$ and $\hat \Psi_{\bf k} = N^{-1/2} \sum_i \hat
\Psi_i$ where $N$ is the number of $k$-points in the sum in
Eq.~(\ref{eq:ham2}).

We need to make an ansatz for ${\bf p_s}({\bf r})$.  For a single
vortex in an isotropic medium, one has ${\bf p_s}({\bf r}) =
(2\pi\hbar/r) {\mathbf{ \hat \theta}}$, where ${\mathbf{ \hat
\theta}}$ is the azimuthal unit vector and the radius $r$ is measured
relative to the centre of the vortex \cite{Vekhter:01}. For the bilayer
model, however, ${\bf p_s}({\bf r})$ is not isotropic: the chains
provide a conduction channel along the ${\bf \hat y}$ direction which
is in parallel with the isotropic plane conduction channel.  We mimic
this anisotropy by assuming that the superfluid momentum will be
similar to that of a single-layer superconductor with an anisotropic
(diagonal) effective mass tensor ${\bf M}$ with $M_{yy} < M_{xx}$.
(For the single-layer model, we take $M_{xx}=M_{yy}$.)  We then have
two requirements which must be satisfied: $\nabla \times {\bf p_s} =
2\pi \hbar \sum_{\bf R} \delta^2({\bf r}-{\bf R})$ and $\nabla \cdot
{\bf v_s} \equiv \nabla \cdot {\bf M}^{-1} \cdot {\bf p_s} =0$.  The
first requirement introduces vortex cores at the vortex lattice sites
${\bf R}$, while the latter incompressibility requirement is strictly
true in regions where $\Delta({\bf r})$ is uniform.  This pair of
equations is solved by
\begin{equation}
{\bf p_{s}} ({\bf r}) = \frac{2\pi\hbar}{L^2} 
{\sum_{\bf G}}^\prime e^{i{\bf G}\cdot{\bf r}} 
\frac{i {\bf M} \cdot ({\bf G}\times {\bf \hat z})}{M_{xx}G_y^2 + M_{yy}G_x^2} 
\label{eq:p}
\end{equation}
where $\sum^\prime$ indicates that ${\bf G}=0$ is excluded from the
sum, ${\bf G}$ are reciprocal lattice vectors of the magnetic unit
cell (we assume a square lattice here) with area $L^2$ and magnetic
length $L$.
The results do not depend strongly on the ratio $M_{yy}/M_{xx}$, which
we take to be 0.6 for the parameters chosen above.  This choice
minimizes $\nabla \cdot {\bf j}({\bf r})$, where ${\bf j}({\bf r})$ is
the total (plane and chain) current in the bilayer,
\begin{equation}
{\bf j}({\bf r}) = \frac{1}{N}\sum_{\bf k} \sum_{n=1}^2 \langle {\bf
v}_{1{\bf k}} + {\bf v}_{2\bf k} \rangle_{\bf r},
\end{equation}
and $\langle \ldots\rangle_{\bf r}$ indicates the expectation value
with respect to $\hat H({\bf r})$, Eq.~(\ref{eq:ham2}).  In principle,
one could improve on the approximation of Eq.~(\ref{eq:p}) by
determining ${\bf p_s}({\bf r})$ self-consistently from ${\bf j}({\bf
r})$; however this will not change the qualitative physics of the
vortex core expansion.

We then solve self-consistently for the order parameter
\begin{equation}
\Delta({\bf r}) = -\frac{V}{N} \sum_{\bf k} [\cos(k_xa)-\cos(k_ya)] 
\langle \psi_{1-{\bf k}\downarrow} \psi_{1{\bf k}\uparrow}\rangle_{\bf r},
\end{equation}
with $V=1.7$.  Self-consistent solutions find that $\Delta({\bf r})$
vanishes near the vortex core centre and obtains an asymptotic value
$\Delta_\mathrm{max} = 0.35$ far from the vortex core.   In order to
measure the vortex core size, we define a quantity $\delta \Delta({\bf
r}) = \Delta_\mathrm{max} - \Delta({\bf r})$.  The vortex core size is
then defined by the first moment of the radial coordinate $r$ with
respect to $\delta \Delta({\bf r})$:
\begin{equation}
R = \frac{\sum_{\bf r} r\delta\Delta({\bf r})}{\sum_{\bf r} \delta
\Delta({\bf r})},
\end{equation}
where $r=0$ corresponds to the vortex core centre.  For presentation
purposes, $R$ is shown relative to the BCS coherence length $\xi_0
\equiv \hbar v_F/\pi\Delta_\mathrm{max}$, where $v_{F}$ is the average
of the Fermi velocity on the Fermi surface.  The magnetic field is
related to the magnetic length by $H = \Phi_0/L^2$ where $\Phi_0$ is
the superconducting flux quantum.  For presentation purposes, $H$ is
scaled by the upper critical field, $H_{c2} \equiv
\Phi_0/2\pi\xi_0^2$, so that $H/H_{c2} = 2\pi\xi_0^2/L^2$.
  

\end{document}